\documentclass[aps,preprint,prd,nofootinbib]{revtex4}

\usepackage{amsmath}
\usepackage{graphicx}
\RequirePackage[colorlinks,citecolor=blue,urlcolor=blue]{hyperref}
\usepackage{soul}

\usepackage{multirow}
\usepackage{array}

\newcommand{\CLASS}{\texttt{CLASS}}
\newcommand{\fnl}{\tilde{f}_{\rm NL}}
\newcommand{\gnl}{\tilde{g}_{\rm NL}}
\newcommand{\td}{{\rm d}}

\begin{document}

\title{Broad primordial power spectrum and $\mu$-distortion constraints on
primordial black holes}

\author{Zhan-He Wang$^{1,2}$\footnote{\href{wangzhanhe19@mails.ucas.ac.cn}{wangzhanhe19@mails.ucas.ac.cn}}}
\author{Hai-Long Huang$^{1,2}$\footnote{\href{huanghailong18@mails.ucas.ac.cn}{huanghailong18@mails.ucas.ac.cn}}}
\author{Yun-Song Piao$^{1,2,3,4}$\footnote{\href{yspiao@ucas.ac.cn}{yspiao@ucas.ac.cn}}}

\affiliation{$^1$ School of Fundamental Physics and Mathematical
    Sciences, Hangzhou Institute for Advanced Study, UCAS, Hangzhou
    310024, China}

\affiliation{$^2$ School of Physical Sciences, University of
Chinese Academy of Sciences, Beijing 100049, China}

\affiliation{$^3$ International Center for Theoretical Physics
    Asia-Pacific, Beijing/Hangzhou, China}

\affiliation{$^4$ Institute of Theoretical Physics, Chinese
    Academy of Sciences, P.O. Box 2735, Beijing 100190, China}



\begin{abstract}

Supermassive black holes (SMBHs) might originate from supermassive
primordial black holes (PBHs). However, the hypothesis that these
PBHs formed through the enhancement of the primordial curvature
perturbations has consistently faced significant challenges due to
the stringent constraints imposed by $\mu$-distortion in the
cosmic microwave background (CMB). In this work, we investigate
the impact of non-Gaussianity on $\mu$-distortion constraint in
the context of broad power spectra. Our results show that, under
the assumption of non-Gaussian curvature perturbations, a broad
power spectrum may lead to weaker $\mu$-distortion constraints
compared to the Gaussian cases. Our findings highlight the
potential of the broad power spectrum to alleviate the
$\mu$-distortion constraints on supermassive PBHs under large
non-Gaussianity.

\end{abstract}

\maketitle


\section{Introduction}


Primordial black holes (PBHs), first proposed by Refs.
\cite{Hawking:1971ei,Carr:1974nx,Zeldovich:1967lct}, could have
formed in the early Universe from the collapse of high-density
regions due to primordial fluctuations. Then the PBHs has been a
subject of extensive research in cosmology, with implications for
dark matter (DM)
\cite{PhysRevD.94.083504,Chapline:1975ojl,Meszaros:1975ef,Carr:2020gox,Calza:2024fzo,Calza:2024xdh},
and supermassive black holes (SMBH) residing in galactic nuclei
\cite{Carr:2023tpt,Nakama:2016kfq}, see also recent reviews
\cite{Carr:2016drx,Garcia-Bellido:2017fdg,Sasaki_2018,annurev:/content/journals/10.1146/annurev-nucl-050520-125911,Carr:2021bzv,Green_2021,2022arXiv221105767E,2024arXiv240406151P,Domènech_2024,Khlopov:2008qy}.
More recently, it has been showed that the merger of supermassive
PBHs (SMPBHs) ($10^{6-10}M_\odot$)\footnote{In principle, the mass
range of a PBH can extend from $10^{-18}M_\odot$ to
$10^{16}M_\odot$. Those smaller than $10^{-18}M_\odot$ would have
evaporated by now due to Hawking radiation, or see recent
\cite{Calza:2024fzo,Calza:2024xdh}.} might be also the source of
nano-Hertz gravitational wave background,
e.g.\cite{Huang:2023chx,Huang:2023mwy,Hooper:2023nnl,Gouttenoire:2023nzr,
Depta:2023qst,Guo:2024iye}.
The supermassive black holes (SMBHs) observed at high redshifts by
the James Webb Space Telescope (JWST) are posing a challenge to
the $\Lambda$CDM model
\cite{Volonteri:2021sfo,Dayal:2024zwq,Hai-LongHuang:2024vvz,Hai-LongHuang:2024gtx,Huang:2024aog,Guo:2024iye}.

However, the scenario of SMPBHs is strongly constrained by
spectral distortions in the CMB due to an enhanced dissipation of
acoustic waves. Specifically, although the CMB photon spectrum is
expected to follow a perfect blackbody radiation spectrum
\cite{Fixsen:1996nj,Mather:1993ij,Mather:1990tfx}, small
deviations can be caused by injecting extra energy into the photon
bath or altering the photon number density
\cite{Zeldovich:1969ff,Chluba:2011hw,illarionov1975comptonization,sunyaev1970interaction}.
The corresponding mechanisms include the dissipation of acoustic
waves, the adiabatic cooling of electrons, the injection of energy
through Hawking evaporation and accretion of PBHs in the early
universe, and the Sunyaev-Zeldovich effect of galaxy clusters in
the late universe
\cite{Chluba:2016bvg,DeZotti:2015awh,sunyaev1972observations,Birkinshaw:1998qp}.
According to different redshift intervals, the deviations can be
classified into $y$ ($z \lesssim 5 \times 10^4$) and $\mu$ ($ 5
\times 10^4  \lesssim z \lesssim 2 \times 10^6$) -distortions. The
redshift corresponding to the PBHs within the SMPBH mass range
falls within the redshift range of $\mu$-distortion. Thus,
$\mu$-distortion sets stringent constraints on the amplitude of
the primordial power spectrum at the corresponding scales. It has
been found that for the monochromatic primordial power spectrum,
the maximal PBH mass permitted by the constraints of
$\mu$-distortion is approximately $10^4M_{\odot}$
\cite{Kohri:2014lza}, while for the broad power spectrum the
constraints will be strengthened \cite{Gow:2020bzo,Pritchard:2024vix}.

Currently, there are two main roads to avoid the $\mu$-distortion
constraints: one is to consider a new PBH mechanism, different
from large primordial perturbations that leads to collapse
\cite{Huang:2023chx,Huang:2023mwy,Kasai:2024tgu,Ai:2024cka}\footnote{In
a landscape with multiple anti-de Sitter (AdS) vacua, AdS bubbles
can collapse into PBHs \cite{Lin:2021ubu,Lin:2022ygd}, which might
be accompanied by a stochastic gravitational wave background,
e.g.\cite{Li:2020cjj}. Thus it is also worth exploring whether AdS
vacua in early universe,
e.g.\cite{Ye:2020btb,Jiang:2021bab,Ye:2021iwa,Wang:2024dka,Wang:2024hwd},
can be relevant to the generation of SMPBHs. }, and the other is
to consider large non-Gaussianity of primordial perturbations
\cite{Nakama:2017xvq,Unal:2020mts,Gow:2022jfb,Hooper:2023nnl,Sharma:2024img,Cai:2023uhc,Balaji:2024wkv}.
In this paper, we still focus on the latter. Non-Gaussianity in
the primordial power spectrum
can significantly enhance the probability to form PBHs
due to enhanced tail distributions
\cite{Byrnes:2012yx}. Therefore, it plays a crucial role in
setting the abundance of PBHs in various mass ranges, directly
impacting their viability as DM candidates and as potential seeds
for SMBHs \cite{Young:2013oia,Franciolini:2018vbk}\footnote{The
non-Gaussianities have also a close relationship to PBH clustering
leading to the potential to be tested by observation
\cite{Hai-LongHuang:2024kye}.}. The $\mu$-distortion constraints
on monochromatic power spectrum with non-Gaussianities have been
studied in \cite{Nakama:2017xvq,Hooper:2023nnl,Sharma:2024img}.

However, the $\mu$-distortion constraint on broad power spectrum
in the context of non-Gaussianity is still rare.
In this paper, we investigate the $\mu$-distortion constraint on
non-Gaussian broad log-normal (LN) and broken power law (BPL)
power spectrum,
which allow flexibility in modelling different inflationary
scenarios \cite{Lan:2024gnv}, using the latest PBH abundance
calculation method
\cite{Young:2022phe,Ferrante:2022mui,Gow:2022jfb}.
Unlike the monochromatic spectrum, where non-Gaussian effects
amplify $\mu$-distortion constraints in large $k$ space, we find
that broad power spectrum can exhibit a weaker non-Gaussian
$\mu$-distortion constraint (specifically the quadratic local
non-Gaussianity) relative to Gaussian cases, especially as the
spectrum is broader. This suggests that the interplay between
spectral width and non-Gaussianity might have unexpected
implications for the constraints of SMPBHs and requires further
investigation.

This paper is outlined as follows. In \autoref{sec:Formalism and
bounds}, we review the methods used to calculate PBH abundance and
$\mu$-distortion constraints, by which we calculate and present
$\mu$-distortion constraints for quadratic local non-Gaussianity
using the log-normal power spectrum in \autoref{sec:log-normal}.
In \autoref{sec:BPL}  we extend this calculation to the BPL power
spectrum. We conclude in \autoref{sec:conclusions} and our main
results can be seen in \autoref{tab:main_result}.

\begin{table}[]
    \centering
    \begin{tabular}{c|c|c|c|c|c}
    \hline
     & \multirow{2}{*}{$\delta$-function} & \multicolumn{2}{c|}{LN} & \multicolumn{2}{c}{BPL} \\
     \cline{3-4} \cline{5-6} & & $\Delta=0.3$ & $\Delta=1$ & $(4,10,1)$ & $(4,3,20)$ \\
    \hline
    Gaussian &  4.25&  3.83&  2.55& 3.67& 2.63\\
    \hline
    NG $(\kappa =0.1)$ &  4.28& 3.98& 2.63& 3.83& 2.77\\
    NG $(\kappa =1)$ &  4.31& 4.22& 3.54& 4.07& 3.42\\
    NG $(\chi^2)$&  4.47& 4.25& 4.00& 4.01& 3.67\\
    \hline
    \end{tabular}
    \caption{The maximum PBH mass, $\log \frac{M_{\rm PBH}}{M_\odot}$, allowed by $\mu$-distortion under the assumption $f_{\rm PBH}=1$ for different types of primordial power spectrum: $\delta$-function \eqref{eq:delta}, log-normal
    \eqref{eq:lognormal} and broken power law \eqref{eq:BPL} form.
    We discuss the scenarios where primordial spectrum peaks
    obeying Gaussian and non-Gaussian statistics under the assumption of local (perturbative or not) non-Gaussianity.
    The parameter $\kappa$ represents the strength of perturbative non-Gaussianity, defined as $\fnl \sqrt{ A_{\rm G}}$, and $\chi^2$ represents non-perturbative non-Gaussianity. It can be seen from the table that
    the enhancement of non-Gaussianity
    relaxes the $\mu$-distortion constraints on the maximum PBH mass, and
    this trend becomes more pronounced as the power spectrum broadening
    increases. Thus, non-perturbative non-Gaussianity reduces the impact
    of power spectrum width on $\mu$-distortion constraints.}
    \label{tab:main_result}
\end{table}

\section{Methods}
\label{sec:Formalism and bounds}


\subsection{PBH abundance calculation}\label{subsection:PBH-abundance}

In this subsection, we present the calculation of PBH abundance
for broad power spectrum in the non-Gaussian case, following
\cite{Ferrante:2022mui}. A PBH form when a sufficiently large
primordial perturbation reentry the event horizon. The compaction
function $\mathcal{C}$
\cite{Musco:2018rwt,Young:2019osy,Young:2019yug} is the most
appropriate parameter for assessing whether the perturbation will
collapse to form a PBH, and it is defined as
\begin{equation}
    \mathcal{C}(r,t) \equiv 2 \frac{\delta M(r,t)}{R(r,t)},
\end{equation}
where $\delta M(r,t)$ is the mass excess within a sphere areal radius $R(r,t)$, $r$ is the radius coordinate in the spherical coordinate.
And the resulting PBH mass follows a critical scaling law
\cite{Musco:2008hv,Musco:2012au}
\begin{equation}
    M_{\rm PBH} =  \mathcal{K} M_k (\cal C -\cal C_{\rm th}\rm )^\gamma,
\end{equation}
where
\begin{equation}
    M_k\simeq 17\left(\frac{g}{10.75}\right)^{-1/6}
    \left(\frac{k}{10^6\, {\rm Mpc}^{-1}}\right)^{-2} M_\odot
    \label{M-k-relation}
\end{equation}
is the Hubble horizon mass corresponding to a comoving scale $k$.
Here, $g$ is the number of relativistic degrees of freedom which
equal to 10.75. The threshold $\cal C_{\rm th}$ can be estimated
following \cite{Musco:2020jjb} for different shapes of power
spectrum. Furthermore, we fix $\cal K  \rm = 4.4$ and $\gamma =
0.38$ \cite{Iovino:2024tyg}.

In the non-Gaussian case,  in which the locally parameterization is $\mathcal {R} =  F (\mathcal {R} _ {\rm G})$ , the compaction function can be characterized via its Gaussian component $\cal C_{\rm G}$ as follows
\begin{equation}
    \cal C = \cal C_{\rm l} - \rm \frac{1}{4\Phi}  \cal C_{\rm l}^{\mathrm{2}},~~~~\rm with~~\cal C_{\rm l} = \cal C_{\rm {G} }\frac{\td \mathnormal{F}}{\td \cal R_{\rm G}},
\end{equation}
where $\Phi = \frac{2}{3}$ during radiation dominated period.  $\cal C_{\rm G}$ is related to Gaussian primordial perturbation $\cal R_{\rm G}$ via
\begin{equation}
    \mathcal{C}_{\mathrm {G}} (r) = -2\Phi r  \frac{\td \cal R_{\rm G}}{\td r}.
\end{equation}
The mass fraction $\beta_k(M_{\rm PBH})$ can be calculated as follows:
\begin{equation}
    \beta_k(M_{\rm PBH}) = \int_{\cal D}\td \mathcal{C} \frac{M_{\rm PBH}}{M_k} P_{\rm G}({\cal C}_{\rm G}, {\cal R}_{\rm G}) \td \cal C_{\rm G} \td \cal R_{\rm G},
\end{equation}
where $\cal D = \{\cal C\rm(\cal C_{\rm G}  ,\cal R_{\rm G} \rm)  >  \cal C_{\rm {th}}  \wedge  \cal C\rm_{\rm l}(\cal C_{\rm G}  ,\cal R_{\rm G} \rm)  <  2 \Phi      \}$.  The two-dimensional Gaussian distribution is
\begin{equation}
    P_{\rm G}(\mathcal {C}_{\rm G},\mathcal {R}_{\rm G}) = \frac{1}{2\pi \sigma_c\sigma_r\sqrt{1-\gamma^2_{cr}}}\mathrm{exp}\left[-\frac{\mathcal {R} _{\rm G}^2}{2\sigma_r^2}-\frac{1}{2(1-\gamma^2_{cr})}\left( \frac{\mathcal {C}_{\mathrm {G}}}{\sigma_c}-\frac{\gamma_{cr}\mathcal {R}_{\rm G}}{\sigma_r} \right)^2\right].
\end{equation}
The correlators are given by
\begin{align}
    & \sigma_c^2=\frac{4 \Phi^2}{9} \int_0^{\infty} \frac{\mathrm{d} k}{k}\left(k r_m\right)^4 W^2\left(k, r_m\right)T^2(k,r_m)\mathcal {P_R}(k)
    \,, \\
    &\sigma_{c r}^2=\frac{2 \Phi}{3} \! \int_0^{\infty} \!\! \frac{\mathrm{d} k}{k} \!\left(k r_m\right)^2~\!W\!\!\left(k, r_m\right) ~\!W_s\!\left(k, r_m\right)~ \!T^2(k,r_m)\mathcal {P_R}(k)
    \,, \\
    &\sigma_r^2=\int_0^{\infty} \frac{\mathrm{d}k}{k} W_s^2\left(k, r_m\right)T^2(k,r_m)\mathcal {P_R}(k) \,,
\end{align}
with $\gamma_{cr} \equiv \frac{\sigma^2_{cr}}{\sigma_c\sigma_r}$ and $r_m \simeq   2.4 \times 10^{-7} {\rm Mpc}~\left(\frac{g}{10.75}\right)^{1/12}
  \sqrt{ \frac{ M_H}{ M_\odot}}$. The top-hat window function $W(k,r_m)$, spherical-shell window function $W_s(k,r_m)$ and the transfer function $T(k,r_m)$  can be found in \cite{Young:2022phe}.
The PBH mass function can be obtained directly from the mass fraction:
\begin{equation}
     \frac{\mathrm {d} f_{\rm PBH}}{\mathrm {d \ln} M_{\rm PBH}}
= \frac{1}{\Omega_{\rm DM}}\int \frac{\mathrm{d} M_k}{M_k} \, \beta_k(M_{\rm PBH} ) \left(\frac{M_{\rm eq}}{M_k}\right)^{1/2} \!\!,
\end{equation}
where  $M_{\rm eq}\simeq 2.7\times10^{17}M_\odot$ is the horizon mass at radiation-matter equality and $\Omega _{\rm DM}=0.12 h^{-2}$ is the cold dark matter density.  The PBH abundance $f_{\rm PBH} = \frac{\rho_{\rm PBH}}{\rho_{\rm DM}} \Big|_{\rm today}$can be expressed after integration:
\begin{align}\label{eq:df_PBH_th}
    f_{\rm PBH} =
    &
\frac{1}{\Omega_{\rm DM}}
\int \frac{\td M_{\rm PBH}}{M_{\rm PBH}}
\int_{M_{k}^{\rm min}}
    \frac{\td M_{k}}{M_{k}} \left(\frac{M_{\rm eq}}{M_{k}}\right)^{1/2}
        \left[  1-\frac{\mathcal{C}_{\rm th}}{\Phi}-\frac{1}{\Phi}\left( \frac{M_{\rm PBH}}{\mathcal{K}M_k}\right)^{1/\gamma}\right]^{-1/2}\nonumber\\
    &\times \frac{{\cal K}}{\gamma} \left(\frac{M_{\rm PBH}}{\mathcal{K} M_{k}}\right)^\frac{1+\gamma}{\gamma}
    \int \td\mathcal{R}_{{\rm G}} P_{\rm G}({\cal C}_{\rm G}(M_{\rm PBH},\mathcal{R}_{\rm G}), \mathcal{R}_{\rm G}) \left(\frac{dF}{d\mathcal{R}_{\rm G}}\right)^{-1}\!\!\!\!.
\end{align}


\subsection{$\mu$-distortion calculation}

For the $\mu$-distortion limits that constrain the amplitude of
perturbations that form SMBHs, we
employ the numerical techniques detailed in Refs.
\cite{Cyr:2023pgw,Tagliazucchi:2023dai,Sharma:2024img}
to calculate the power spectrum amplitude that saturates the COBE-FIRAS
$\mu$-type distortion limit.
To be concentrate, it can be quantified by the
dimensionless quantity $\mu$ and calculated as follows:
\begin{equation}
    \mu = \int dk \frac{k^2}{2\pi^2}P_{\mathcal R}(k)W_{\mu}(k)=\int \frac{dk}{k}\mathcal{P}_{\mathcal R}(k)W_{\mu}(k),
    \label{eq:mu-calculate}
\end{equation}
where $P_{\mathcal R}(k)$ is the curvature power spectrum and
$\mathcal{P}_{\mathcal R}(k)$ is the dimensionless curvature power
spectrum and they are related by
\begin{equation}
    \mathcal{P}_{\mathcal R}(k)=\frac{k^3}{2\pi^2} P_{\mathcal R}(k).
    \label{eq:P_R-P_R}
\end{equation}
$W_{\mu}(k)$ is the window function which contains details regarding
the $\mu$-distortion. It can be numerically calculated by \CLASS{} and
we adopt the results from \cite{Sharma:2024img}:
\begin{equation}
    W_\mu^{\rm}(k)= \exp\left[\sum_{n=0}^6 z_n \ln^n \left(\frac{k}{1\,
    {\rm Mpc}^{-1}}\right)\right]~,
    \label{eq:W_F}
    \end{equation}
with coefficients $z_n$ provided in the Table 1 of \cite{Sharma:2024img}.
We utilize the updated value of
$\mu<4.7\times10^{-5}$ from Bianchini and Fabbian \cite{Bianchini:2022dqh},
which is a factor of 2 more stringent than the constraint originally
reported by the COBE collaboration \cite{Fixsen:1996nj} and slightly
tighter than the TRIS result \cite{Gervasi:2008eb}.


\begin{figure}[htbp]
    \centering
    \includegraphics[width=0.67\textwidth]{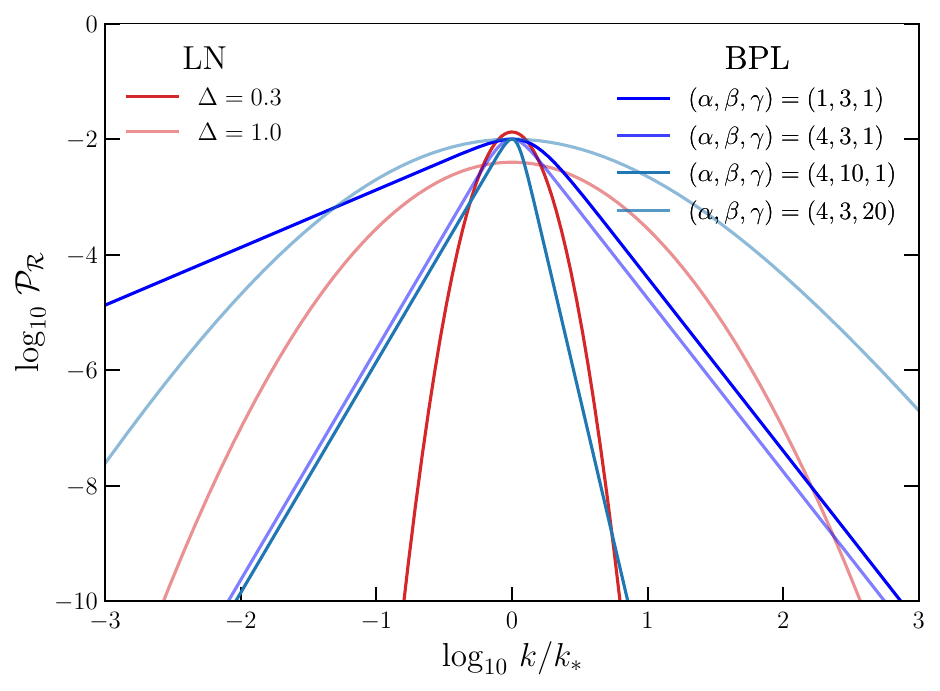}
    \caption{The primordial broad power spectrum against the comoving-to-peak wavenumber ratio with log-normal (red, \eqref{eq:lognormal}) and
    broken power law (blue, \eqref{eq:BPL}) form, where $A_{\rm LN}=A_{\rm BPL}=10^{-2}$.}
    \label{fig:power_spectrum}
\end{figure}

\section{Log-normal power spectrum case }\label{sec:log-normal}

A broad primordial power spectrum is more likely to arise in
realistic inflation models.
A commonly used broad power spectrum is the log-normal form given by
\cite{Gow:2020bzo,Ferrante:2022mui}
\begin{equation}
    \mathcal{P} _{\mathcal{R}}(k) = \frac{A_{\rm LN}}{\sqrt{2\pi}\Delta}
    \exp\left(-\frac{\ln^2\left(k/k_{*}\right)}{2\Delta^2}\right),
    \label{eq:lognormal}
\end{equation}
where $A_{\rm LN}$ is the amplitude, $k_{*}$ is the peaked
wavenumber and $\Delta$ is the width. We recover the
$\delta$-function spectrum
\begin{equation}\label{eq:delta}
   \mathcal{P}_{\mathcal{R}}(k)=A_{\rm LN}~k_*\delta(k-k_*)
   \end{equation}
for $\Delta \rightarrow 0$. Another form of the broad power
spectrum, the broken power law, will be discussed in
\autoref{sec:BPL}.
Then we show the impact of
broadening on $\mu$-distortion under both Gaussian and
non-Gaussian conditions.

\subsection{Gaussian case}

\begin{figure}[htbp]
    \centering
    \includegraphics[width=0.67\textwidth]{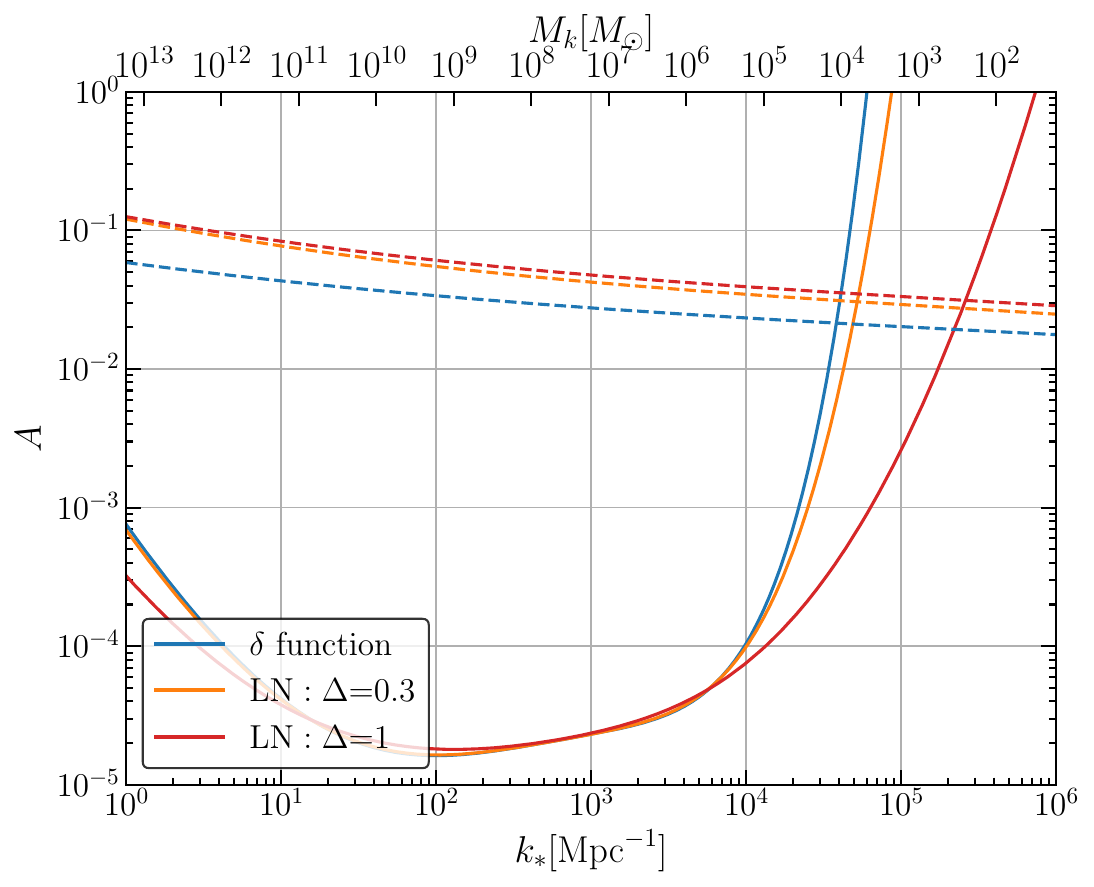}
    \caption{The $\mu$ constraints on the variance assuming Gaussian curvature perturbations for the $\delta$-function power spectrum \eqref{eq:delta} and log-normal power spectrum \eqref{eq:lognormal} respectively.
    The dashed lines show the corresponding PBH constraints (we choose $f_{\rm PBH}=1$ constraints throughout the paper) on the variance calculated by \eqref{eq:df_PBH_th}.
    The upper x-axis shows the horizon mass in
    solar mass units (which is approximately the PBH mass) corresponding to each $k_{\star}$ value.}
    \label{fig:Gaussian_case_delta_lognormal}
\end{figure}

In the Gaussian case, the variance of the curvature is given by
\begin{equation}
    A = \langle \mathcal{R} ^2(\vec{x}) \rangle = \int_{0}^{\infty}
    \frac{\td k}{k}\mathcal{P} _{\mathcal{R}}(k)=A_{\rm LN}.
\end{equation}
We note that the width $\Delta$ does not affect the variance. Substituting
\eqref{eq:lognormal} into \eqref{eq:mu-calculate} yields the
$\mu$-distortion constraint on the variance of the curvature for the
log-normal power spectrum,
as shown by the solid lines
in \autoref{fig:Gaussian_case_delta_lognormal}. As a contrast, we
also plot the constraints for the $\delta$-function power
spectrum \eqref{eq:delta}.

As the width of log-normal power spectrum $\Delta$ is broader, the
$\mu$-distortion constraints become increasingly stronger at large
$k$. The reason is explained as follows. When $k$ takes values
between $10$ and $10^4~\mathrm{Mpc}^{-1}$, the window function
$W_{\mu}$ is relatively large (due to the fact that the
$\mu$-distortion primarily arises during this period). As the
width of the power spectrum is broader, scales further away from
$k_{*}$ acquire greater amplitudes. Therefore, when
$k_{*}>10^4~\mathrm{Mpc}^{-1}$, the contribution of the power
spectrum to the $\mu$-distortion increases according to
\eqref{eq:mu-calculate}, leading to a stricter constraint on $A$.
Furthermore, the amplitude $A$ required for $f_{\rm PBH }=1$
slightly increases with the width. Then, the maximum
PBH mass $\mu$-distortion required decreases which is smaller than
$10^4~M_{\odot}$, in agreement with the assumptions presented in
the Refs.\cite{Gow:2020bzo,Byrnes:2024vjt}.

\subsection{Non-Gaussian case}

Non-Gaussianity has a significant impact on the properties of
PBHs. Firstly, PBHs formed from non-Gaussian perturbations tends
to cluster. Observational constraints on such clustering may limit
the ability to circumvent $\mu$-distortion constraints for the
generation of SMPBHs through large non-Gaussianity
\cite{Shinohara:2021psq,DeLuca:2022uvz,Shinohara:2023wjd,Khlopov:2004sc}.
Secondly, there are stringent constraints on any non-Gaussian
correlation between PBH-forming scales and CMB scales, through the
restrictions imposed by photon-dark matter isocurvature
perturbations \cite{Tada:2015noa,Young:2015kda,vanLaak:2023ppj}.
More discussion can refer to \cite{Byrnes:2024vjt}.

Primordial local non-Gaussianity can be characterised by a Taylor
expansion of comoving curvature perturbation
\cite{Young:2013oia}
\begin{equation}
    {\cal R}(\vec{x})=
        {\cal R}_{\rm G}(\vec{x})
        + \fnl
        \left( {\cal R}_{\rm G}^2(\vec{x})
        - \langle{\cal R}^2_{\rm G}(\vec{x})\rangle
        \right) + \gnl
         {\cal R}_{\rm G}^3(\vec{x})
         + \cdots,
        \label{eq:Rg}
    \end{equation}
where $\fnl\equiv 3f_{\rm NL}/5 $ and $\gnl\equiv 9g_{\rm NL}/25$ are
non-linearity parameters quantifying the magnitude of non-Gaussianity.
Here, we focus only on the quadratic term and $\fnl>0$ because it
corresponds to an improvement of the production of PBH and leave higher order
terms for further research.
Additionally, we assume that the Gaussian field $\cal R_{\rm G}$ has a log-normal form
for its power spectrum, with a peak at $k_{\star}$  and amplitude $A_{\rm G}$.

Then we examine two different types of non-Gaussianity. One is the
perturbative non-Gaussianity, i.e. the linear term is predominant,
specifically $\fnl \sqrt{A_{\rm G}} \ll 1$. Another is
non-perturbative non-Gaussianity, i.e. either $\fnl$ or even
higher-order terms dominate to such an extent that the power
spectrum is entirely governed by these non-linear terms.

\subsubsection{Perturbative non-Gaussian constraints}

Here we present constraints in both the strictly perturbative limit
of $\fnl \sqrt{ A_{\rm G}} \ll 1$ and the borderline case of perturbativity
$\fnl \sqrt {A_{\rm G}}= 1$. Using the contraction that $\langle {\cal R}_{\rm G}^4\rangle
=3\langle {\cal R}_{\rm G}^2\rangle^2=3A_{\rm G}^2$, \eqref{eq:Rg} demonstrates
that the total variance is given by
\begin{equation}
    A \equiv \langle \mathcal{R} ^2 \rangle = A_{\rm G} + 2 \tilde{f}_{\rm NL}^2 A_{\rm G}^2.
\end{equation}
If we define $\kappa$ as $\fnl \sqrt{ A_{\rm G}}$, then the variance can be expressed by
\begin{equation}
    A = A_{\rm G}(1+2\kappa^2).
\end{equation}

To calculate $\mu$-distortion in the non-Gaussian case, we need to
focus on the primordial power spectrum. As discussed in
\cite{Byrnes:2007tm}, corrections to this power spectrum arise from
a single one-loop diagram, which leads to
\begin{equation}
    P_{\cal R}(k) = P_{\mathcal R_{\rm G}}(k)+\tilde{f}^2_{\rm{NL}}
    P_{\mathcal{R}^2_{\rm G}}(k) = P_{\mathcal R_{\rm G}}(k)+
    2\tilde{f}^2_{\rm{NL}} \int \frac{d^3q}{(2\pi)^3}P_{
    \mathcal{R}_{\rm G}}(q)P_{\mathcal{R}_{\rm G}}(|\vec{k}-\vec{q}|).
    \label{eq:P_R_loop}
\end{equation}
Substituting \eqref{eq:P_R_loop} into \eqref{eq:mu-calculate}, we
obtain the spectral distortion
\begin{equation}
    \mu
        = \int \frac{\td k}{k} \mathcal{P} _{\mathcal{R}_{\rm G}}(k)
        W_{\mu}(k)+\int \frac{\td k}{k} \tilde{f}^2_{\rm NL} \mathcal{P}
        _{\mathcal{R}^2_{\rm G}}(k) W_{\mu}(k).
\end{equation}
To calculate $\mathcal{P} _{\mathcal{R}^2_{\rm G}}(k)$, we use
spherical coordinates as $(q,\theta,\phi)$ assuming that $\vec{k}$
points in $z$ direction which leads to
\begin{multline}
    \mathcal{P} _{\mathcal{R}^2_{\rm G}}(k) = \int_{-1}^{1}\td y \int_{0}^{\infty}\td q
    \frac{{A_{\rm G}}^2}{\Delta^2} \frac{1}{2\pi}
    \frac{k^2}{q~(\sqrt{k^2-2kqy+q^2})^3} \\
    \times \exp\left[
    -\frac{1}{2\Delta^2} \Bigg(
    \ln^2\frac{q}{k_{*}}
    + \ln^2\frac{\sqrt{k^2-2kqy+q^2}}{k_{*}}
    \Bigg)\right].
    \label{eq:P_R2}
\end{multline}
where $y\equiv \mathrm{cos} \theta$. We solve \eqref{eq:P_R2} numerically.

\begin{figure}[htbp]
    \centering
    \includegraphics[width=0.67\textwidth]{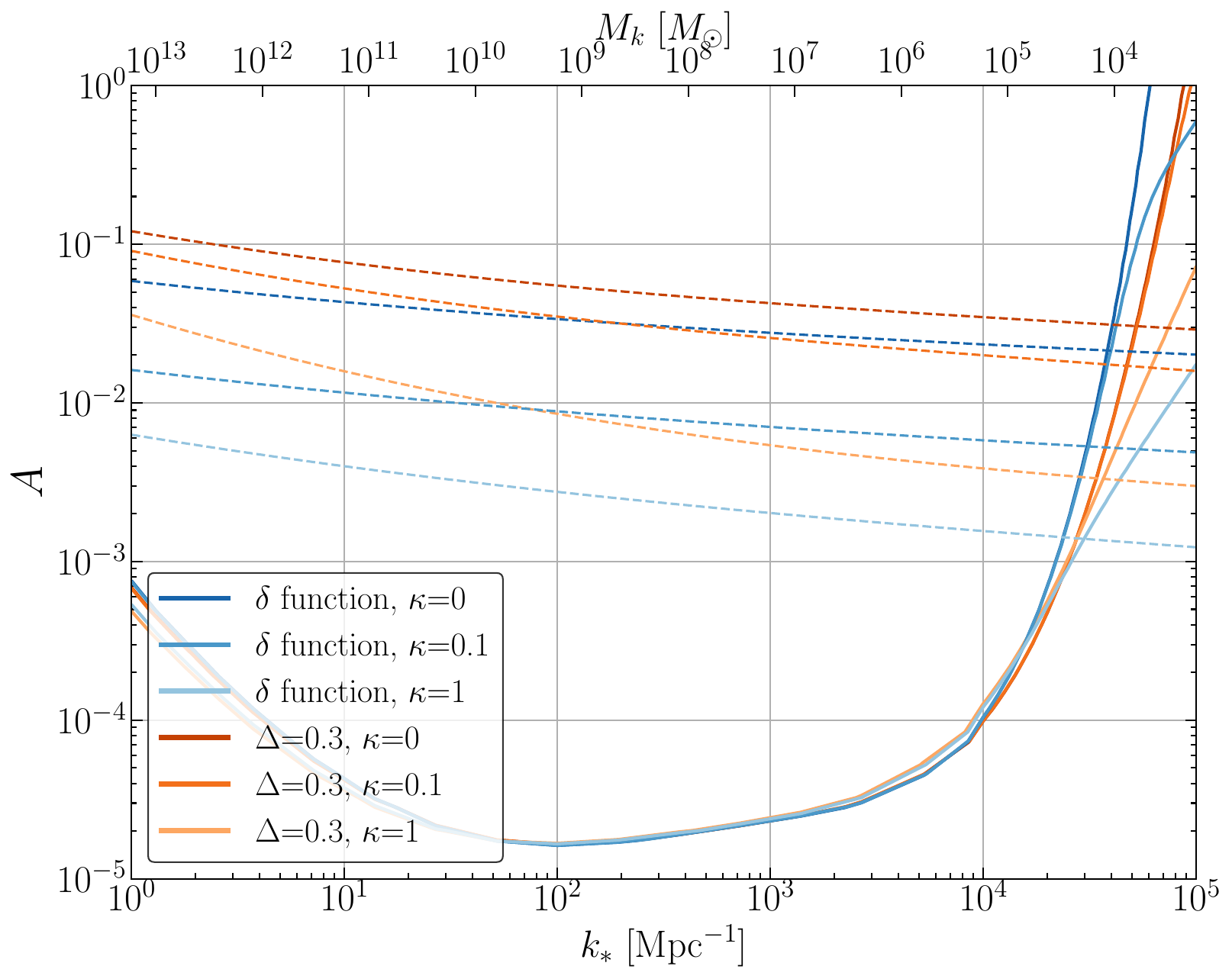}
    \caption{Constraints on the variance assuming perturbative non-Gaussian curvature perturbations, assuming log-normal power spectrum \eqref{eq:lognormal} with $\Delta=0.3$ and
    $\delta$-function power spectrum \eqref{eq:delta} respectively. The solid lines represent the $\mu$-distortion constraints, while the dashed lines represent the abundance constraints requiring $f_{\rm PBH}=1$. Here, $\kappa$ is defined as $\fnl \sqrt{ A_{\rm G}}$ and quantifies the relative non-Gaussian contribution to the total variance.}
    \label{fig:NG_case_for_kappa_delta_0.3}
\end{figure}

The method for determining the PBH abundance for arbitrary values
of $\fnl$ is described in \autoref{subsection:PBH-abundance}. We note that
non-Gaussianity would relax the PBH abundance constraints on the variance $A$.
\autoref{fig:NG_case_for_kappa_delta_0.3}  illustrates the perturbative
non-Gaussian scenarios for log-normal power spectrum with $\Delta=0.3$
compared with $\delta$ function case.
We find that, under strict perturbative conditions, the $\mu$-distortion
nearly coincides with the Gaussian case with broadening power spectrum.
Examining the case at the perturbation boundary, we find that broadening
weakens the $\mu$-distortion constraints. However, due to the difference in
abundance constraints on $A$, the maximum PBH mass in the $\delta$ function
case is still larger.

\subsubsection{Non-perturbative case :$\chi^2$ distribution}

\begin{figure}[htbp]
    \centering
    \includegraphics[width=0.67\textwidth]{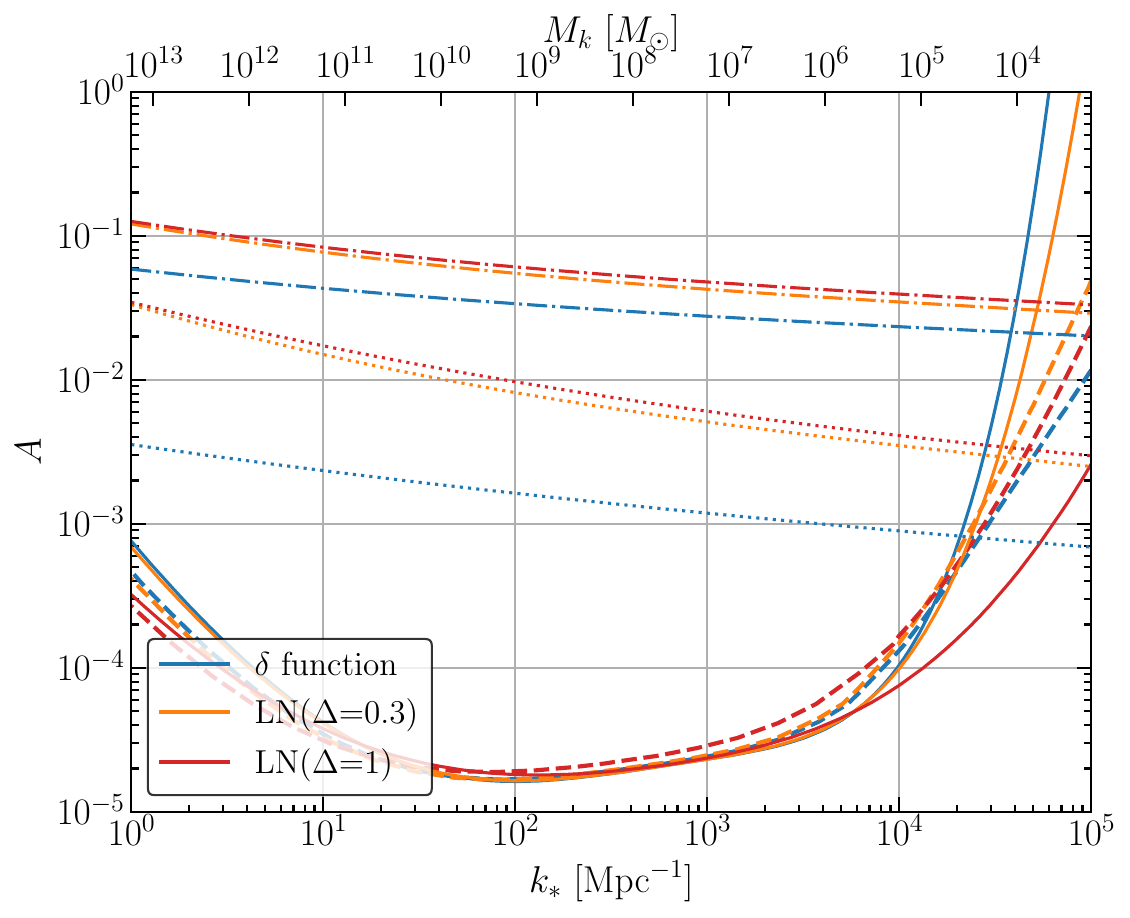}
    \caption{Constraints on the variance for
    Gaussian (solid) and $\chi^2$ (dashed) statistics for
    $\delta$-function \eqref{eq:delta} and LN \eqref{eq:lognormal} power spectrum with different $\Delta$. The horizontal dash-dotted line and dotted line represent the $f_{\rm PBH}=1$ constraints for PBHs in the Gaussian and $\chi^2$ cases, respectively.  }
    \label{fig:NG_case_for_delta}
\end{figure}

In this case, the comoving curvature perturbation can be expressed in
the following form in terms of its Gaussian component
\begin{equation}
    \mathcal{R}(\vec{x}) = \mathcal{R}^2_{\rm G}(\vec{x}) - \langle{\cal R}
    ^2_{\rm G}(\vec{x})\rangle.
\end{equation}
Then the variance can be given by
\begin{equation}
    A \equiv \int_{0}^{\infty}\frac{\td k}{k}\mathcal{P} _{\mathcal{R}^2_{\rm G}}(k)=2A^2_{\rm G}.
\end{equation}
The $\mu$-distortion can then be computed using \eqref{eq:mu-calculate}
with $\mathcal{P_{R}}(k)$ substituted with \eqref{eq:P_R2}. The result
can be put in the form

\begin{equation}
    \mu = \int \frac{\td k}{k} \mathcal{P} _{\mathcal{R}^2_{\rm G}}(k) W_{\mu}(k).
\end{equation}
\autoref{fig:NG_case_for_delta} illustrate the $\chi^2$
non-Gaussian scenario compared with Gaussian scenario for
different width $\Delta$. It is clear that as the width of the
power spectrum broadens, the $\mu$-distortion constraint in the
Gaussian case gradually strengthens in the large $k$ region. In
contrast, the $\mu$-distortion constraint in the non-Gaussian case
exhibits a slight decrease followed by an increase. This results
in the non-Gaussian case having a stronger constraint in the large
$k$ region than Gaussian case when the power spectrum width is
small, but a weaker constraint when the width is larger.

We can explain this result by considering the effect of power
spectrum width $\Delta$ on $\mathcal{P_{R_{\rm G}}}(k)$
\eqref{eq:lognormal} and $\mathcal{P} _{\mathcal{R}^2_{\rm G}}(k)$
\eqref{eq:P_R2}. When the width $\Delta$ is small, $\frac1k
\mathcal{P} _{\mathcal{R}^2_{\rm G}}(k)$ is larger than $\frac1k
\mathcal{P_{R_{\rm G}}}(k)$ in the region below $k_{*}$, which
results in a greater contribution to the $\mu$-distortion and
stronger constraint on the variance $A$ when $k_{*}\gtrsim 10^4
\rm{Mpc^{-1}}$. However, as the width $\Delta$ braodens, this
relationship reverses. In the above discussion, we have considered
the impact of the $\chi^2$ distribution on the variance $A$.

\section{Broken power law power spectrum case}\label{sec:BPL}

The broken power law (BPL) form of power spectrum is defined as
\begin{equation}
    \mathcal{P}_{\mathcal{R}}(k) = A_{\rm BPL} \frac{(\alpha+\beta)^{\gamma}}{
    \left[
    \beta(k/k_{*})^{-\alpha/\gamma}+\alpha(k/k_{*})^{\beta/\gamma}\right]^{\gamma}}
    \label{eq:BPL}
\end{equation}
where $\alpha$ describes the growth, $\beta$ describes the decay and
$\gamma$ the width of the spectrum at the scale $k_{*}$. This
parametrisation can describe a common class of spectra that exhibit peaks,
often associated with single-field inflation or curvaton models, with typical values for $\alpha$ being $0 < \alpha \lesssim 4$.
Additionally, in quasi-inflection-point models that give rise to stellar-mass PBHs, one typically expects $\beta \gtrsim 0.5$, while
for curvaton models $\beta \gtrsim 2$.

To generate the parameter space constraint plots for the BPL equivalent
to the log-normal case, we fixed two of the parameters $(\alpha,\beta,\gamma)$.
The fixed values were identified $(\alpha,\beta,\gamma)=(4,3,1)$ \cite{Franciolini:2023pbf,Pritchard:2024vix}.
The parameter choices for comparison can be seen in \autoref{fig:power_spectrum}.
It is important to emphasize that the width of the BPL power spectrum
broadens as $\alpha$ decreases, $\beta$ decreases, or $\gamma$ increases.
The procedure for calculating the $\mu$-distortion and PBH abundance using
the BPL power spectrum in (non-)Gaussian case is similar to that for the log-normal case,
see \autoref{sec:Formalism and bounds}.

\subsection{Gaussian case}

The variance now is given by
\begin{equation}
    A = \langle \mathcal{R} ^2(\vec{x}) \rangle = \int \frac{\td k}{k} \mathcal{P}_{\mathcal{R}}(k)
        = A_{\rm BPL} \int \frac{\td k}{k} \frac{(\alpha+\beta)^{\gamma}}{
    \left[
    \beta(k/k_{*})^{-\alpha/\gamma}+\alpha(k/k_{*})^{\beta/\gamma}\right]^{\gamma}}.
    \label{eq:A_G_BPL}
\end{equation}
The expression above cannot be solved analytically. We emphasize that,
in contrast to the log-normal case, the variance $A$ depends on the
parameters $\alpha$, $\beta$ and $\gamma$ of the BPL power spectrum besides
the amplitude.

\begin{figure}[htbp]
    \centering
    \includegraphics[width=0.67\textwidth]{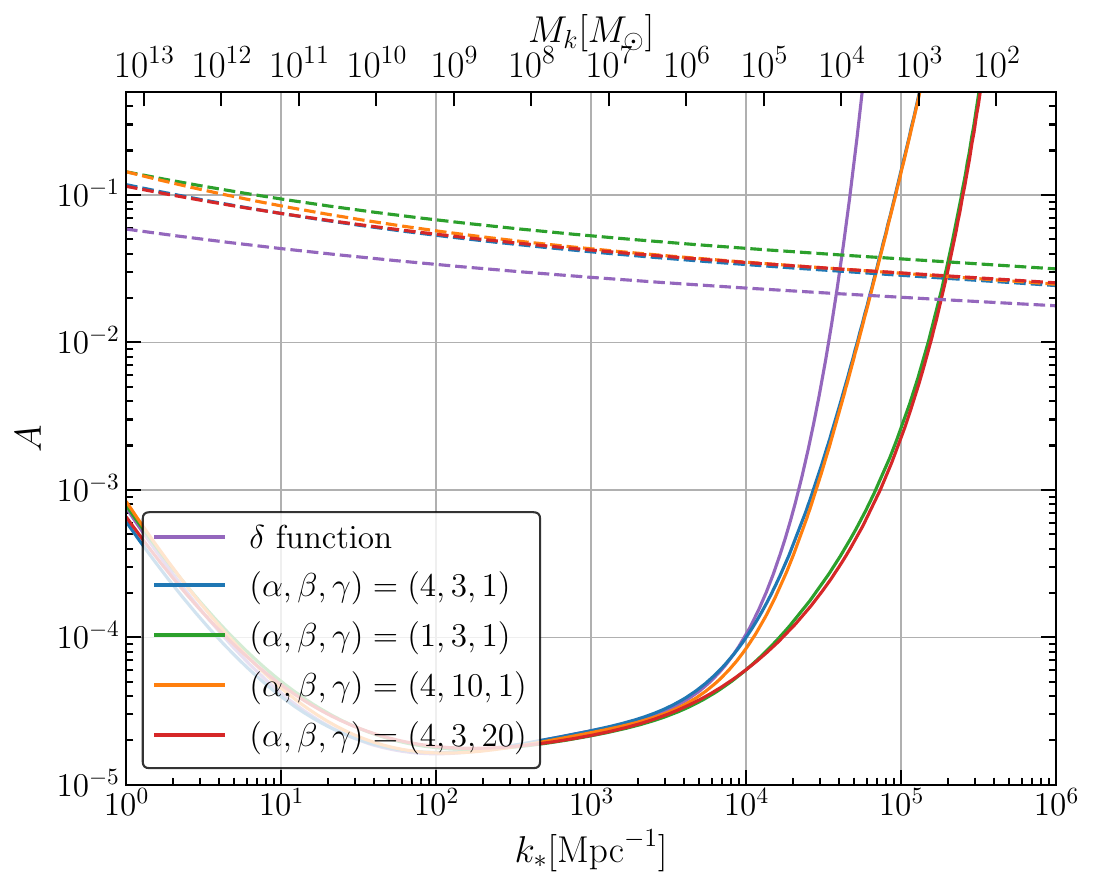}
    \caption{Constraints on the variance assuming Gaussian curvature perturbations
    for BPL \eqref{eq:BPL} compared with $\delta$-function \eqref{eq:delta} power spectrum. We can see that only when the width of the BPL power spectrum undergo a significant change, the $\mu$-constraints show noticeable variations in the high-$k_{\star}$ region .}
    \label{fig:Gaussian_case_BPL}
\end{figure}

The constraints of different parameter combinations on
$\mu$-distortion and PBH abundance are summarized in \autoref{fig:Gaussian_case_BPL}.
We can observe that when the broadening of the BPL power spectrum
undergoes significant changes, the $\mu$-distortion constraints show noticeable
variations. Moreover, the broader the spectrum, the stronger the constraints
in the high-$k_{\star}$ region, while in the low-$k_{\star}$ region, the constraints are similar
to those in the log-normal case.  In contrast to the log-normal case, the changes
in abundance constraints due to different parameter combinations are minimal.
In summary, the broader the spectrum, the smaller the allowed maximum PBH mass.

\subsection{Non-Gaussian case}
\subsubsection{Perturbative non-Gaussian constraints}

\begin{figure}[htbp]
    \centering
    \includegraphics[width=0.67\textwidth]{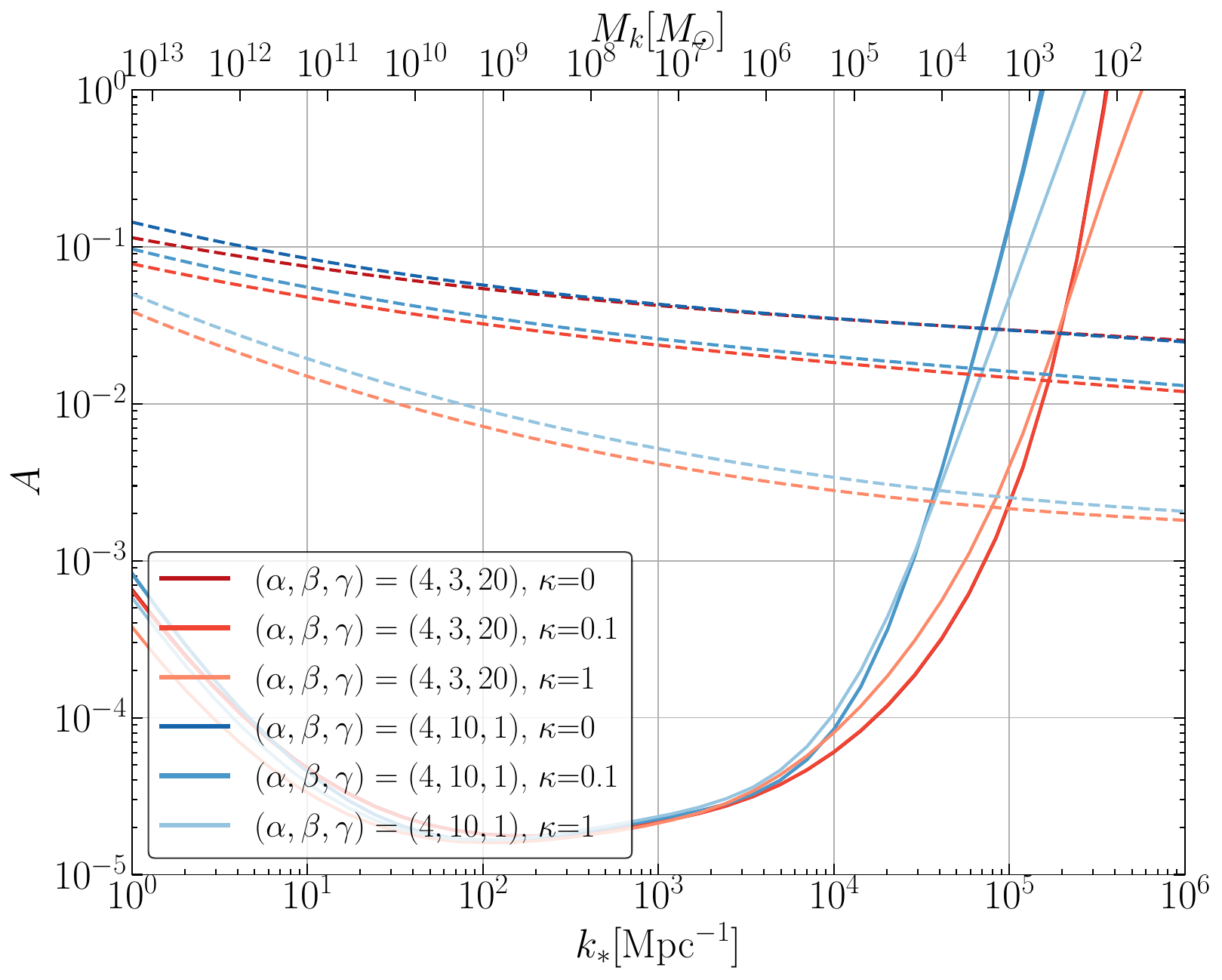}
    \caption{The solid (dashed) lines represent the $\mu$-distortion ($f_{\rm PBH}=1$ abundance) constraints on the variance. $\kappa$ is defined as $\fnl \sqrt{ A_{\rm G}}$ and quantifies the relative non-Gaussian contribution to the total variance. We note that the $\mu$-distortion constraints in the Gaussian case nearly coincide with those for $\kappa = 0.1$.}

    \label{fig:BPL_NG_perturbative}
\end{figure}

Similar to the log-normal case, we first need to
calculate $P_{\mathcal{R}^2_{\rm G}}(k)$.
However, the result now is more complex:
\begin{align}
    P_{\mathcal{R}^2_{\rm G}}(k) ={}& 2 \int \frac{d^3q}{(2\pi)^3} P_{\mathcal{R}_{\rm G}}(q) P_{\mathcal{R}_{\rm G}}(|\vec{k}-\vec{q}|) \nonumber \\
    ={}& 2 \int \frac{d^3q}{(2\pi)^3} \frac{2\pi^2}{q^3} A_{\rm BPL} \frac{(\alpha+\beta)^{\gamma}}{\left [\beta\left(\frac{q}{k_{*}}\right)^{-\frac{\alpha}{\gamma}}+\alpha\left(\frac{q}{k_{*}}\right)^{\frac{\beta}{\gamma}}\right ]^{\gamma}}
     \frac{2\pi^2}{|\vec{k}-\vec{q}|^3} A_{\rm BPL} \frac{(\alpha+\beta)^{\gamma}}{\left [\beta\left(\frac{|\vec{k}-\vec{q}|}{k_{*}}\right)^{-\frac{\alpha}{\gamma}}+\alpha\left(\frac{|\vec{k}-\vec{q}|}{k_{*}}\right)^{\frac{\beta}{\gamma}}\right ]^{\gamma}} \nonumber \\
    ={}& 2 \int_{0}^{\infty} dq \int_{-1}^{1} dy \; \pi^2 \frac{A^2_{\rm BPL}}{q} \frac{1}{(\sqrt{k^2-2kqy+q^2})^3}
     \frac{(\alpha+\beta)^{\gamma}}{\left [\beta\left(\frac{q}{k_{*}}\right)^{-\frac{\alpha}{\gamma}}+\alpha\left(\frac{q}{k_{*}}\right)^{\frac{\beta}{\gamma}}\right ]^{\gamma}} \nonumber \\
    &\times \frac{(\alpha+\beta)^{\gamma}}{\left [\beta\left(\frac{\sqrt{k^2-2kqy+q^2}}{k_{*}}\right)^{-\frac{\alpha}{\gamma}}+\alpha\left(\frac{\sqrt{k^2-2kqy+q^2}}{k_{*}}\right)^{\frac{\beta}{\gamma}}\right ]^{\gamma}}. \label{eq:P2_BPL}
\end{align}
The result definitely cannot be solved analytically. \autoref{fig:BPL_NG_perturbative}  illustrates the perturbative non-Gaussian scenarios for the cases $(\alpha,\beta,\gamma) = (4,3,20)$ and $(4,10,1)$.  In the perturbative non-Gaussian region, the $\mu$-distortion constraints for case $(4,3,20)$ are always stronger than those for case $(4,10,1)$.

\subsubsection{Non-perturbative case: $\chi^2$ distribution}

\begin{figure}[htbp]
    \centering
    \includegraphics[width=0.67\textwidth]{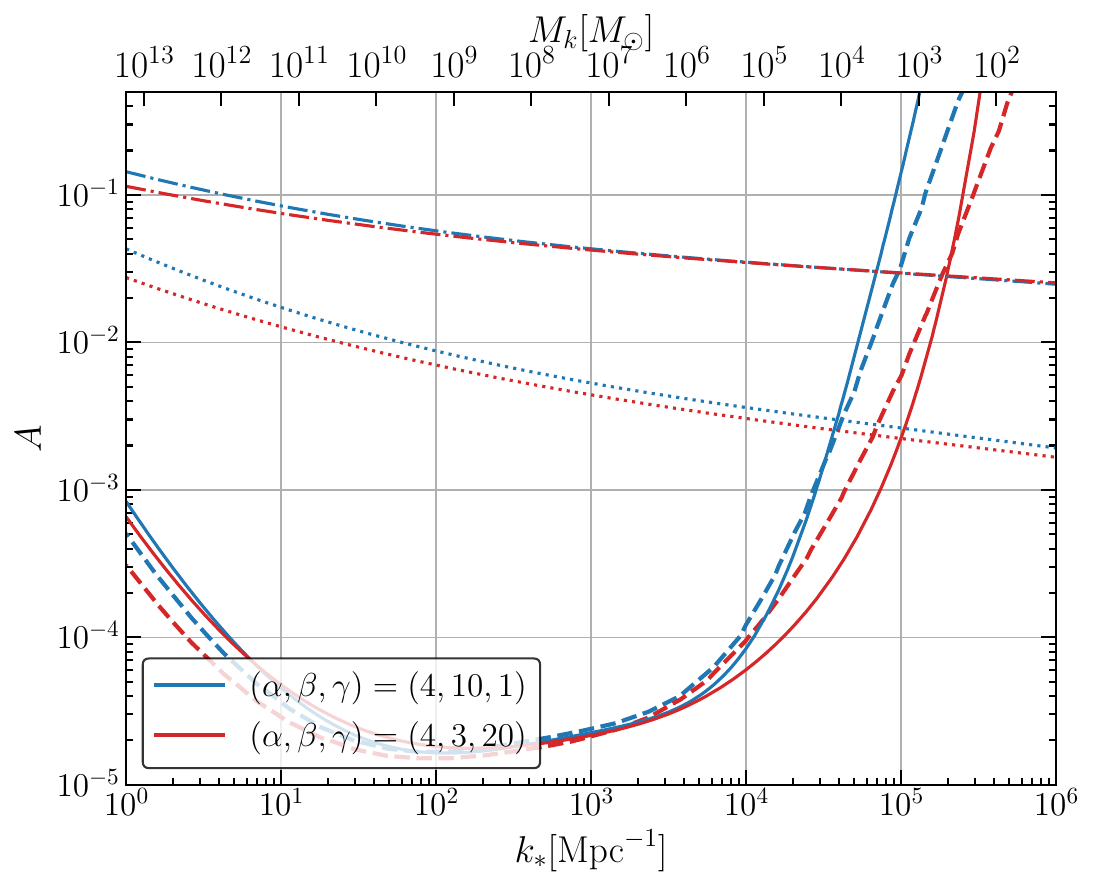}
    \caption{Constraints on the variance for
    Gaussian (solid) and $\chi^2$ (dashed) statistics for different $(\alpha,\beta,\gamma)$. The horizontal dash-dotted line and dotted line represent the $f_{\rm PBH}=1$ constraints for PBHs in the Gaussian and $\chi^2$ cases, respectively. }
    \label{fig:BPL_NG_chi}
\end{figure}

To obtain the constraints in the $\chi^2$ case, we follow a procedure
similar to the log-normal scenario.
By substituting
\eqref{eq:P2_BPL} and \eqref{eq:P_R-P_R} into \eqref{eq:mu-calculate}
and using \eqref{eq:A_G_BPL}, we obtain \autoref{fig:BPL_NG_chi}.

The commonality among these two parameter combinations is that
the $\mu$-distortion constraints in the $\chi^2$ case are not significantly different
from those in the Gaussian case. Due to the impact of non-Gaussianity on
the PBH abundance calculation, considering the $\chi^2$ non-Gaussian form,
the allowed maximum PBH mass is increased by less than an order of
magnitude compared to the Gaussian case. However, the specific value
remains below $10^5 M_{\odot}$.

\section{Conclusions}\label{sec:conclusions}

In recent years, $\mu$-distortion constraints have been widely
applied to probe the viability of SMPBHs originating in the early
universe. Prior studies have established $\mu$-distortion as a
stringent observational limit on the primordial power spectrum and
introduce non-Gaussianity to evade or relax the constraint.
However, these studies largely assumed $\delta$-function power
spectrum \eqref{eq:delta} and do not capture the effects of a more
realistic power spectrum with certain width. Our study addresses
this gap by investigating $\mu$-distortion constraints for broad
power spectrum models, specifically log-normal
\eqref{eq:lognormal} and broken power law \eqref{eq:BPL}
distributions, both of which reflect potential early-universe
scenarios more accurately than single-peaked models.

For the log-normal case, when the width is small ($\Delta \lesssim
0.3$), the situation is essentially the same as the monochromatic
case, as expected \cite{Byrnes:2024vjt,Gow:2020bzo}. However, as
the width is broader, the effects of the width and non-Gaussianity
on the distortion constraints do not simply add together, when the
width reaches a certain threshold, the non-Gaussian constraints
may even become weaker than those in the Gaussian case in the
large $k$ space. For the BPL power spectrum, the $\mu$-distortion
constraints in the small $k$ space is similar to log-normal and
$\delta$-function case, regardless of the existence of
non-Gaussianity. The broadening of the power spectrum is
positively correlated with the $\mu$-distortion constraints in the
high-$k_{\star}$ region, while non-Gaussianity enhances the
constraints in the case of narrow broadening and weakens them in
the case of wide broadening, the same as log-normal case.

For both broad power spectrum above, we found that
$\mu$-distortion limits restrict the maximum PBH mass to
approximately $10^4 M_{\odot}$, suggesting that PBHs with masses
sufficient to seed SMBHs (i.e.$\gtrsim 10^6 M_{\odot} $) cannot
form under the current constraints (see \autoref{tab:main_result}
for the summary). However, if higher-order non-Gaussianity is
considered as suggested in \cite{Byrnes:2024vjt}, we expect the
horizontal line representing the PBH abundance to bypass the
$\mu$-distortion constraints at the bottom in the
$k_{\star}$-range from $10^2$ to $10^4~ \rm Mpc^{-1}$. Our
analysis shows that the $\mu$-distortion constraints in this
region on the variance are independent of both the broadening of
the power spectrum and the existence of non-Gaussianity.
Therefore, generating SMPBHs may require higher-order
non-Gaussianity to relax the variance for the abundance of the
corresponding PBHs, a possibility that warrants further
investigation.

\section*{Acknowledgments}

This work is supported by National Key Research and Development
Program of China (Grant No. 2021YFC2203004), NSFC (Grant
No.12075246), and the Fundamental Research Funds for the Central
Universities.

\bibliography{REF}

\end{document}